\documentclass[aps,prl,twocolumn,superscriptaddress,amsmath,amssymb,floatfix]{revtex4-2}
\usepackage{graphicx}
\usepackage{bm}
\usepackage[colorlinks=true,linkcolor=blue,citecolor=blue,urlcolor=blue]{hyperref}

\begin{document}

\title{Nonlocal multipolar spin accumulation in altermagnets beyond d-wave}

\author{J. W. Gonz\'alez}
\affiliation{Departamento de F\'isica, Universidad de Antofagasta, Av. Angamos 601, Casilla 170, Antofagasta, Chile}

\date{\today}

\begin{abstract}
In spin-conserving models of $g$- and $i$-wave altermagnets, the uniform
longitudinal conductivity has no spin-odd component. A point injector
reveals higher moments of the Fermi velocity through the nonlocal spin
accumulation. Boltzmann theory predicts that the rank $m=2,4,6$ of the
first nonzero spin-odd velocity moment sets the leading angular harmonic,
its asymptotic radial dependence $r^{2-m}$, and its amplitude scaling
$\tau^{m-3}$ at fixed distance and injected current. For $g$-wave order,
the fourth moment produces a fourth angular harmonic with an asymptotic
$r^{-2}$ tail. Quantum transport simulations support the radial profiles
predicted by finite-geometry Boltzmann theory without an amplitude fit.
Across the three symmetries, the quantum amplitudes exceed the Boltzmann
predictions based on the Born scattering time by 15--65\% in the windows
analyzed. In a two-contact ribbon, we detect the $g$-wave fourth harmonic at
$6.8\,\sigma$ in an independent set of 1800 disorder realizations using a
preselected radial window. Its amplitude decreases when the altermagnetic
splitting is reduced.
\end{abstract}

\maketitle

Altermagnets combine compensated collinear magnetic order with a nonrelativistic,
momentum-dependent spin splitting of the bands
\cite{Smejkal2022PRX,Smejkal2022Landscape}. Crystal symmetry permits even-parity
splitting patterns of $d$-, $g$-, and $i$-wave character. Spectroscopy has resolved
this splitting in $g$-wave MnTe and CrSb \cite{Krempasky2024,Reimers2024} and in
layered materials with $d$-wave signatures
\cite{Jiang2025KV2Se2O,Zhang2025RbV2Te2O}. 
In earlier work, we explored two-dimensional altermagnetism in Ca$_2$RuO$_4$ monolayers and stacking-engineered MnPS$_3$
bilayers \cite{Gonzalez2025Ca2RuO4,Gonzalez2026MnPS3}. We also predicted
topological chiral magnons in monolayer AgF$_2$ \cite{Gonzalez2026AgF2}
and chirality-split magnons and cycloidal phonons in monolayer CuF$_2$
\cite{leon2026chiral}.

The electrical spin-splitter effect was predicted in compensated magnets \cite{Gonzalez2021splitter}. Lateral spin-valve measurements on Mn$_5$Si$_3$
detect a spin signal that reverses between time-reversed magnetic states,
consistent with the spin-splitter response of a $d$-wave altermagnet
\cite{mencos2026time}. In CrSb films, crystal symmetry engineering
reconstructs the magnetic order, enabling anomalous Hall readout and
electrical switching \cite{Zhou2025CrSb}. These experiments motivate a
search for electrical signatures of higher-wave order.
In the spin-conserving models considered here, the uniform longitudinal
conductivity is governed by the second moment of the Fermi velocity.
For $g$- and $i$-wave order, symmetry makes this moment identical in the
two spin channels, so the spin-odd part of the uniform longitudinal
conductivity vanishes. The same rank-two limitation applies to transport equations truncated at second order in spatial gradients \cite{Kokkeler2025} and to the local conductivity mechanism for current vortices near point contacts \cite{Herasymchuk2025}. Nonlinear spin currents \cite{Ezawa2025} and equilibrium impurity signatures
\cite{Sukhachov2024Friedel,Hu2025QPI,Petermann2025} provide other probes of
higher-wave symmetry. Can a spatially varying response recover this
information in linear transport?

Here we show that a point injector probes higher moments of the Fermi velocity, revealing them through the spatial harmonics of the nonlocal spin accumulation. The spin-conserving Boltzmann equation predicts a hierarchy of spatial harmonics for $d$-, $g$-, and $i$-wave order. For $g$-wave order, the fourth moment generates a fourth
angular harmonic in the spin accumulation, with an asymptotic $r^{-2}$ tail.
Quantum transport simulations support the radial profile predicted by
Boltzmann theory in finite geometries without an amplitude fit. The fourth
harmonic also remains detectable in a two-contact device adapted from
Ref.~\cite{Herasymchuk2025}.

\begin{figure*}[t]
\includegraphics[width=0.75\textwidth]{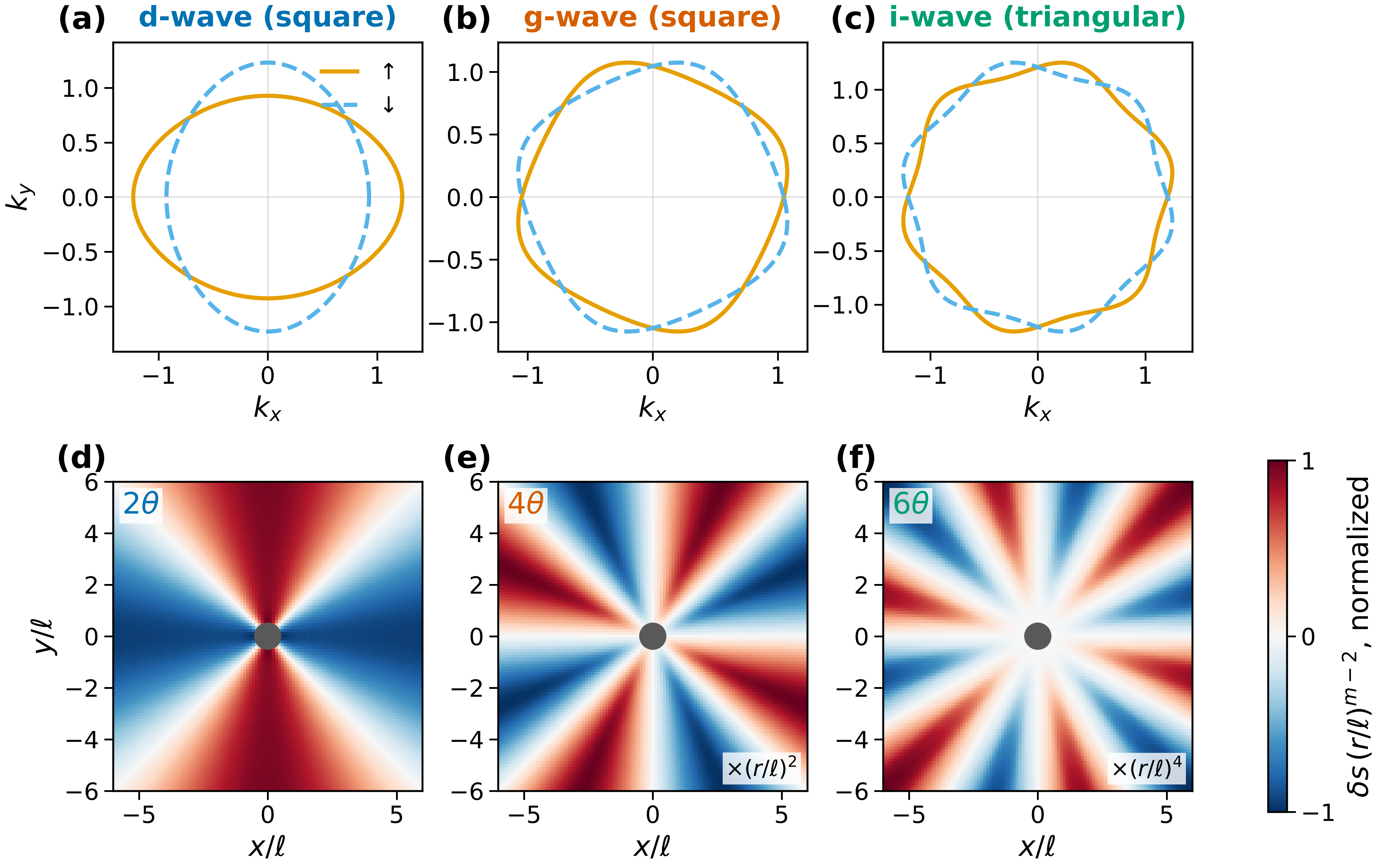}
\caption{(a) to (c) Spin-resolved Fermi surfaces of the $d$-, $g$-, and $i$-wave models (solid: spin up, dashed: spin down). (d) to (f) Spin density $\delta s=n_\uparrow-n_\downarrow$ around a point injector from the BGK
equation (\ref{eq:bgk}) on an infinite plane, multiplied by $(r/\ell)^{m-2}$ and normalized to its maximum
outside $r=\ell/2$ (gray disk); $U=2$ for $d$- and $g$-wave order, $U=3.5$ for $i$-wave order, so that $\ell\approx 9$ lattice spacings.}
\label{fig:maps}
\end{figure*}

\paragraph{\textbf{Models.}}
We use single-orbital tight-binding models with conserved $s_z$.
We label the spin-up and spin-down sectors by $s=+1$ and $s=-1$,
respectively. Their dispersions are
\begin{equation}
E_s(\bm k) = \varepsilon_0(\bm k) + s\,\lambda\,f_X(\bm k).
\label{eq:model}
\end{equation}
Here $\varepsilon_0(\bm k)$ is the spin-independent nearest-neighbor
dispersion and $f_X(\bm k)$ is a symmetry-adapted lattice form factor.
On the square lattice, we use $f_d=\cos k_x-\cos k_y$ and
$f_g=\sin k_x\sin k_y(\cos k_x-\cos k_y)$
\cite{Gondolf2025Local}. On the triangular lattice, we use the
in-plane form factor
$f_i=\prod_{j=1}^{3}\sin(\bm k\cdot\bm a_j)\sin(\bm k\cdot\bm b_j)$
\cite{Ezawa2025}, where $\bm a_j$ and $\bm b_j$ are the nearest- and
next-nearest-neighbor bond vectors, respectively. This product
transforms as the $A_2$ representation of $C_{6v}$.
 We set the zero of energy at the bottom of the spin-independent band
$\varepsilon_0(\bm k)$. The Fermi energy $E_F$ defines the spin-resolved
Fermi contours through $E_s(\bm k)=E_F$. In units of the
nearest-neighbor hopping, we use $(\lambda,E_F)=(0.5,1)$, $(1,1)$,
and $(2,2)$ for $d$-, $g$-, and $i$-wave order, respectively
[Fig.~\ref{fig:maps}(a)--(c)]. Disorder is an on-site potential uniform in $[-U/2,U/2]$, and the Born rate
$\hbar/\tau=2\pi\nu U^2/12$, with $\nu$ the density of states per site and per spin, sets $\tau$.
We define $\ell=v_F\tau$, where $v_F$ is the density-of-states-weighted
Fermi-surface average of $|\bm v|$, averaged over the two spin sectors. For the $g$- and $i$-wave models, symmetry makes the second velocity moment
$\langle v_i v_j\rangle_s$ identical in the two spin sectors, as we verify
numerically. The spin-odd part of the uniform longitudinal conductivity therefore vanishes. To access the spin response in these models, we turn
to the spatially varying density generated by a point injector.

\begin{figure}[t]
\includegraphics[width=0.9\columnwidth]{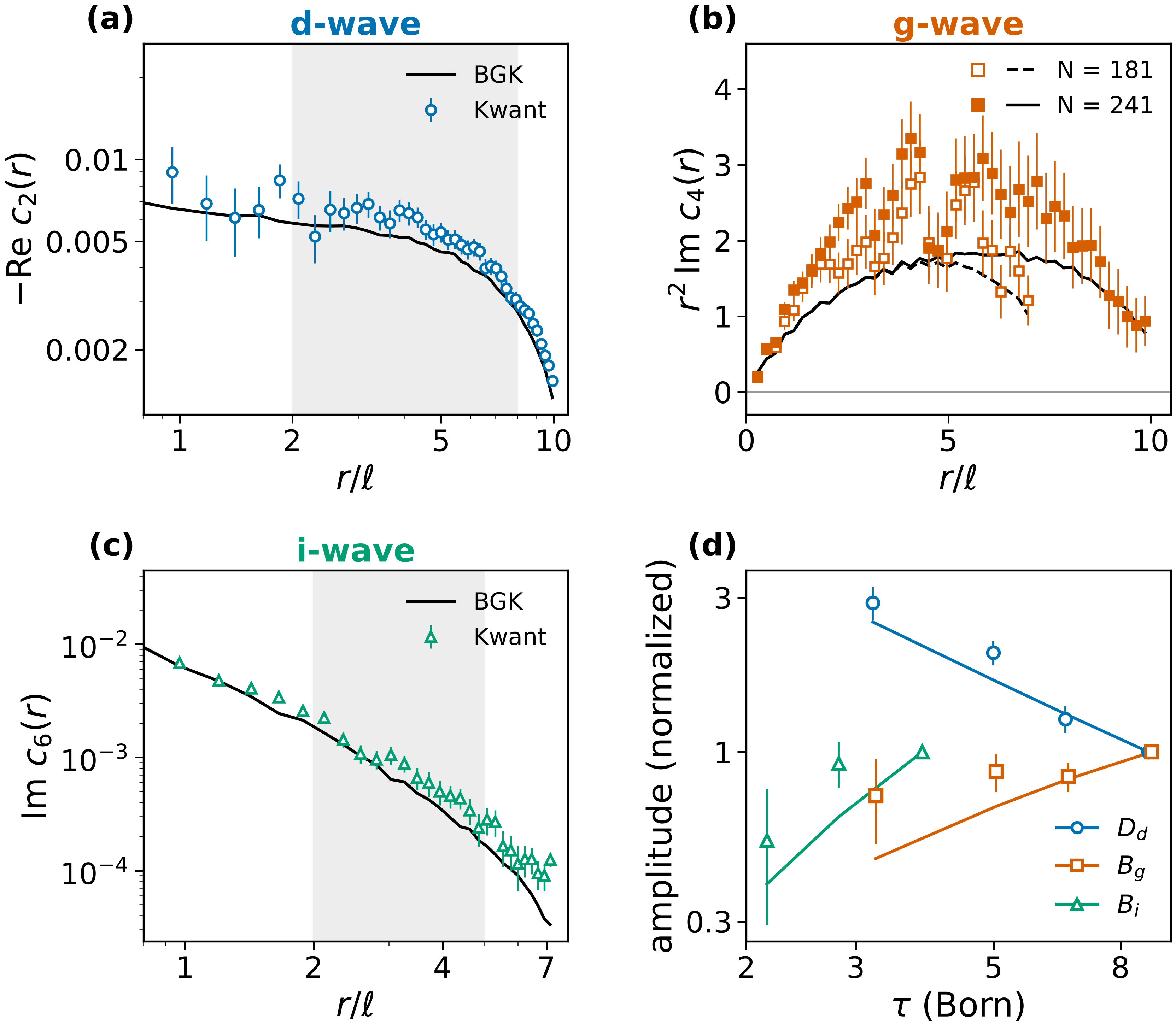}
\caption{Kwant (symbols) against the BGK equation solved in the same finite geometry (lines), absolute
comparison without rescaling. (a) $d$-wave order, $-\mathrm{Re}\,c_2(r)$, 150 realizations; (c) $i$-wave order,
$\mathrm{Im}\,c_6(r)$, 200 realizations; gray bands mark the slope windows. (b) $g$-wave order,
$r^2\,\mathrm{Im}\,c_4(r)$ for $N=181$ (open squares, BGK dashed, 150 realizations) and $N=241$ (filled
squares, BGK solid, 300 realizations). (d) Window averages $D_d$, $B_g$, and $B_i$ of the harmonics weighted by $1$, $r^2$, and $r^4$,
respectively, versus the Born $\tau$, normalized to the weakest disorder. Kwant symbols use the symmetry-fixed
projection with the sign of the clean sample. BGK lines use $|c_m|$. Error bars are bootstrap standard errors
over disorder realizations.}
\label{fig:kwant}
\end{figure}

\paragraph{\textbf{Kinetic theory.}}
We model spin-conserving isotropic scattering with a
Bhatnagar--Gross--Krook (BGK) collision term \cite{BGK1954}
that relaxes each spin distribution toward its own Fermi-surface
average. For a point source of unit injection rate in each spin
sector, isotropic over its Fermi surface, the Fourier component of
the steady-state density is
\begin{equation}
n_s(\bm q)=\frac{\tau A_s(\bm q)}{1-A_s(\bm q)},\qquad
A_s(\bm q)=
\left\langle \frac{1}{1+(\tau\,\bm q\cdot\bm v)^2}\right\rangle_s ,
\label{eq:bgk}
\end{equation}
where $\langle\cdots\rangle_s$ denotes the density-of-states-weighted
Fermi-surface average. Let $q=|\bm q|$ and
$M_{n,s}(\hat{\bm q})=\langle(\hat{\bm q}\cdot\bm v)^n\rangle_s$.
For $q\ell\ll1$, we expand $1-A_s$ through order $q^6$ and invert it
using $n_s=\tau/(1-A_s)-\tau$. This gives
\begin{equation}
\begin{aligned}
n_s(\bm q)={}&
\frac{1}{\tau q^2 M_{2,s}}
+\frac{\tau M_{4,s}}{M_{2,s}^2}-\tau \\
&+\tau^3q^2\left(
\frac{M_{4,s}^2}{M_{2,s}^3}
-\frac{M_{6,s}}{M_{2,s}^2}
\right)+O(q^4).
\end{aligned}
\label{eq:expansion}
\end{equation}
The term $-\tau$ is identical in both spin sectors and cancels in
$\delta s=n_\uparrow-n_\downarrow$.
The spin-odd moments $\delta M_n=M_{n,\uparrow}-M_{n,\downarrow}$
transform under the crystal symmetries as the splitting $f_X(\bm k)$. A rank-$n$ moment contains angular harmonics of order at
most $n$. Thus $\delta M_n=0$ for $n<m$, with $m=2,4,6$ for
$d$-, $g$-, and $i$-wave order, respectively. Subtracting the two
propagators in Eq.~(\ref{eq:bgk}) gives
\begin{equation}
\delta s(\bm q)=
\tau\,\frac{A_\uparrow(\bm q)-A_\downarrow(\bm q)}
{[1-A_\uparrow(\bm q)][1-A_\downarrow(\bm q)]}.
\end{equation}
The numerator $A_\uparrow-A_\downarrow$ first contributes at order
$(\tau q)^m$, while each denominator factor starts at order
$(\tau q)^2$. The leading spin-odd term therefore scales as
$\tau^{m-3}q^{m-4}$, with an angular dependence set by symmetry.
The second-gradient transport terms in
Refs.~\cite{Kokkeler2025,Herasymchuk2025} probe the rank-two response,
whose spin-odd part vanishes for the $g$- and $i$-wave models
considered here.

For $d$-wave order, the first term of Eq.~(\ref{eq:expansion})
produces a leading $\cos 2\theta$ harmonic with amplitude proportional
to $\tau^{-1}$ and independent of $r$ in the infinite plane.
For $g$- and $i$-wave order, we write $M_2$ for the common second
moment of the two spin sectors. For $g$-wave order, the leading
spin-odd term is $\tau\,\delta M_4/M_2^2\propto\sin 4\phi_q$,
where $\phi_q$ is the polar angle of $\bm q$. It has degree zero
in $q$ and retains its angular dependence as $q\to0$.
For $m>0$ and $r>0$, the inverse Fourier transform of $e^{im\phi_q}$ is
$(i^m m/2\pi)e^{im\theta}/r^2$ [Eq.~(S2) of the SM], yielding
$\delta s\propto\tau\sin 4\theta/r^2$.
For $i$-wave order, both $\delta M_2$ and $\delta M_4$ vanish,
and the term $-\tau^3q^2\delta M_6/M_2^2$ yields
$\delta s\propto\tau^3\sin 6\theta/r^4$.

For an isotropic diffusion tensor, the summed density is
$n_0(r)=n_\uparrow(r)+n_\downarrow(r)
\simeq \ln(R/r)/(\pi\tau\bar M_2)$, where $\bar M_2$ is the
spin-averaged second moment and $R$ is set by the drain or sample size.
At fixed band parameters, the resulting relative harmonic amplitude
scales as
\begin{equation}
\frac{\delta s_m(r)}{n_0(r)}
\propto
\left(\frac{\ell}{r}\right)^{m-2}
\frac{1}{\ln(R/r)},
\qquad \ell\ll r\ll R .
\label{eq:relative}
\end{equation}
The absolute harmonics plotted in Fig.~\ref{fig:kwant} are evaluated
without division by $n_0$. Higher angular harmonics allowed by the
same symmetry can share the leading radial dependence. For example,
the $d$-wave response can contain a $\cos 6\theta$ component with the
same $r^0$ dependence as the leading $\cos 2\theta$ term.
Figure~\ref{fig:maps}(d)--(f) shows the full solution of
Eq.~(\ref{eq:bgk}).

\begin{figure}[t]
\includegraphics[width=\columnwidth]{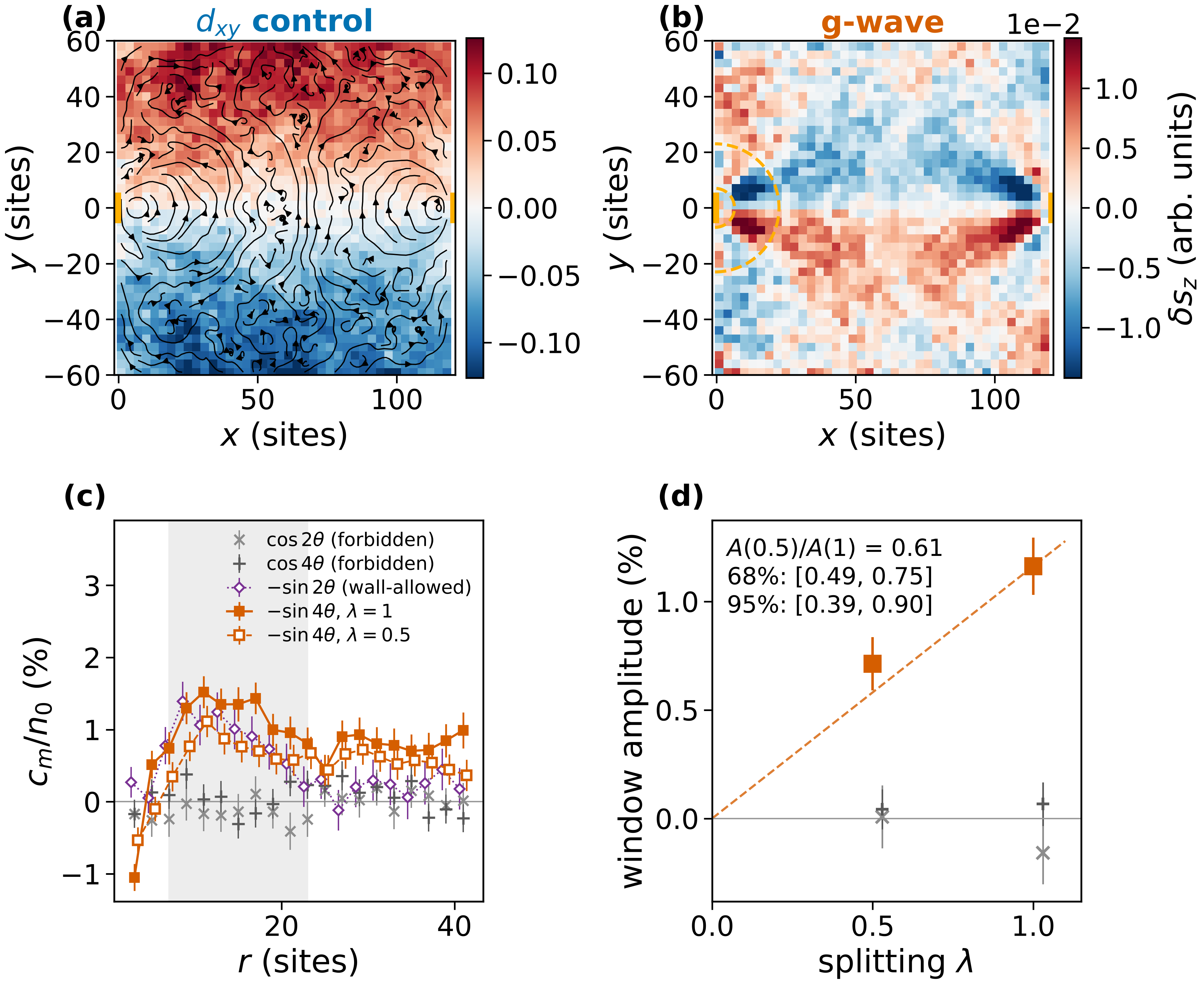}
\caption{Two-contact ribbon (orange bars) as in Ref.~\cite{Herasymchuk2025}. (a) $d_{xy}$ control: spin
density of the states injected from the left contact, averaged over the bias window (arbitrary units), and
spin-current streamlines. (b) $g$-wave model, 2400 realizations; dashed half
rings bound the window of (c) and (d). Both maps are projected on the parity allowed under $y\to-y$.
(c) $g$-wave harmonics on half rings, in percent of the spin-summed left-injected density averaged
over the same half ring. Projections are $-\sin 4\theta$ for
$\lambda=1$ and $0.5$, the wall-allowed $-\sin 2\theta$, and the forbidden $\cos 2\theta$, $\cos 4\theta$
(gray). (d) Window amplitude of $-\sin 4\theta$ versus $\lambda$ (squares), the reference $A(1)\lambda$ (dashed),
and the forbidden harmonics (gray); the paired ratio is given with 68\% and 95\% intervals.}
\label{fig:ribbon}
\end{figure}

\paragraph{\textbf{Quantum transport.}} 
We use Kwant \cite{Kwant2014} to compute the scattering states of the
disordered tight-binding models and obtain the nonequilibrium spin
density. A semi-infinite chain injects electrons at $E_F$ into the
central site of the sample. We use square samples of $N\times N$ sites
with $N=181$ and $241$ for $d$- and $g$-wave order, and a disk of radius
90 lattice spacings for $i$-wave order. A clean absorbing layer
surrounds each sample.

We divide the density in each spin sector by its net injected particle
current. In a clean sample, the integrated outward flux agrees with
the injected current within 2\% [Supplemental Material (SM)~\cite{SM}].
On each ring we compute $c_m(r)=2\langle\delta s\,e^{-im\theta}\rangle_{\rm ring}$
from the normalized spin-density difference. We use the leading
symmetry-allowed harmonic and bootstrap whole disorder realizations. For comparison, we solve the BGK equation
in the corresponding finite domains using the Peierls integral equation
[Eq.~(S1) of the SM~\cite{SM}]. Both calculations give a density per unit injected
particle current, so the comparison requires no amplitude rescaling.
Section~S-I of the SM describes AI assistance with the analysis.

For $g$-wave order [Fig.~\ref{fig:kwant}(b)],
$r^2\,\mathrm{Im}\,c_4$ rises up to $r\approx3\ell$, then varies
more slowly over an intermediate radial range (Sec.~S-IV of the SM). This behavior is
consistent with the $\tau\,\delta M_4$ term in
Eq.~(\ref{eq:expansion}). The $r^2$-weighted profile falls off
at a larger radius in the larger geometry. Averaged over the
same rings ($2\ell<r<7\ell$) and the same 150 disorder
realizations, it increases from $1.98$ at $N=181$ to $2.74$
at $N=241$. The paired increase in Kwant is $0.76$
(95\% confidence interval $0.18$--$1.31$), compared with
$0.16$ in BGK (from $1.50$ to $1.66$).
For $d$-wave order [Fig.~\ref{fig:kwant}(a)], the logarithmic
slope in $2\ell<r<8\ell$ is $-0.54\pm0.13$ in Kwant and
$-0.52$ in BGK. For $i$-wave order [Fig.~\ref{fig:kwant}(c)],
we compare the sixth angular harmonic, $\mathrm{Im}\,c_6$.
Its logarithmic slope in $2\ell<r<5\ell$ is $-2.27\pm0.37$
in Kwant and $-2.59$ in BGK. This agreement supports the
finite-geometry prediction for the sixth harmonic. The
asymptotic $r^{-4}$ law remains a prediction.

Using the Born scattering time, the Kwant/BGK amplitude ratios
are $1.15$ for $d$-wave order, $1.32$ and $1.65$ for $g$-wave
order at $N=181$ and $241$, respectively, and $1.37$ for $i$-wave
order. As a sensitivity test, we repeat the BGK calculation at
$U=2$ with the transport time inferred independently from the
Ohmic conductivity of disordered strips,
$\tau_{\rm eff}\approx0.88\,\tau_{\rm Born}$ (Sec.~S-VI of the SM).
The $d$-wave ratio decreases to $1.02$, while the $g$-wave ratios
increase to $1.46$ and $1.81$ with $\tau_{\rm eff}$ over the same rings
and realizations.
At $N=241$, halving or doubling the absorber strength at fixed
layer width and disorder realizations changes the $g$-wave mean
by less than 5\% of the remaining gap, as bounded by the paired
95\% confidence intervals. Coherent scattering remains a possible
contribution to the discrepancy.

Figure~\ref{fig:kwant}(d) shows the disorder dependence. For each
symmetry, we extract the exponent from Kwant and BGK using the same
fixed radial window (Sec.~S-V of the SM). For $d$-wave order,
$p_d=-1.08\pm0.11$, compared with $-0.91$ in BGK. The estimates $p_g=+0.27\pm0.33$ and $p_i=+1.07\pm0.99$
are compatible with the finite-geometry BGK values of $+0.74$ and
$+1.63$, respectively, and their confidence intervals also
include zero.

\paragraph{\textbf{Two-contact ribbon.}}
Herasymchuk et al.~\cite{Herasymchuk2025} found swirling spin currents
in a square $d$-wave sample with narrow contacts on opposite sides.
In an adapted geometry with $121\times121$ sites and 11-site contacts,
we recover spin-current loops near both contacts and a transverse
spin accumulation [Fig.~\ref{fig:ribbon}(a); Sec.~S-VII of the SM].

In the same geometry, the $g$-wave model at $U=2$ develops spin lobes
at both contacts [Fig.~\ref{fig:ribbon}(b)]. The reference contact
width is comparable to the mean free path, and the fourth harmonic
remains detectable for contact widths of 7, 11, and 15 sites. The
mirror $y\to-y$ exchanges the two spin sectors, so the disorder-averaged
spin accumulation is odd in $y$. The $\cos 2\theta$ and $\cos 4\theta$
projections therefore serve as noise controls, while both
$\sin 2\theta$ and $\sin 4\theta$ are allowed by the device symmetry.
Ballistic contributions from the contacts and walls can have the
same parity.

On half rings around the injector [Fig.~\ref{fig:ribbon}(c)], the
full ensemble of 2400 realizations gives a mean $\sin 4\theta$
amplitude of $-1.16\pm0.13\%$ of the spin-summed left-injected density over
$7\le r\le23$ lattice spacings. The first 600 realizations fixed
this radial window. The remaining 1800 give $-1.04\pm0.15\%$
in the same window, a $6.8\sigma$ signal. In the full ensemble,
the $\sin 2\theta$ component is $-0.87\pm0.20\%$, and the two
forbidden harmonics lie within $1.1\sigma$ of zero.

Halving the altermagnetic splitting gives a paired amplitude ratio
$A(0.5)/A(1)=0.61$ (95\% confidence interval $0.39$--$0.90$)
[Fig.~\ref{fig:ribbon}(d)]. The decrease is consistent with
proportional scaling. The mean fourth harmonic has larger magnitude
at $U=1.5$, while the signal at $U=2.5$ is consistent with zero (Sec.~S-VII of the SM). These controls
establish that the fourth harmonic is detectable without strain
in a two-contact device. The radial profile around the central
injector [Fig.~\ref{fig:kwant}(b)] provides the evidence connecting
this harmonic to the bulk fourth-order term.

\paragraph{\textbf{Experimental signature}}
Ferromagnetic detectors \cite{JohnsonSilsbee1985,Jedema2001} placed
at several angles around a narrow injector could resolve the angular
pattern of the spin accumulation. The leading harmonic changes sign
under angular shifts of $45^\circ$ for $g$-wave order, $90^\circ$
for $d$-wave order, and $30^\circ$ for $i$-wave order. To probe the
$g$- and $i$-wave signals at distances of a few mean free paths,
the injector should be narrower than $\ell$, the detectors should
have sub-$\ell$ spatial resolution, and the spin relaxation length
should exceed the measurement range. Experimental estimates are
given in Sec.~S-VIII of the SM.

\paragraph{\textbf{Conclusion.}}
In a $g$-wave altermagnet, the fourth moment of the Fermi velocity
leaves a spatial signature in the spin accumulation even when the
spin-odd uniform longitudinal conductivity vanishes. Quantum transport
resolves this fourth harmonic and supports its Boltzmann profile in
finite geometries. For $i$-wave order, Boltzmann theory predicts a
sixth harmonic with an asymptotic $r^{-4}$ tail. The angular phase
tracks the orientation of the nodal planes, and reversing the order
at fixed crystal orientation reverses the signal. These two signatures
could allow an array of spin-sensitive contacts to determine the
nodal orientation and distinguish opposite domains electrically.

\paragraph{\textbf{AI use disclosure.}}
The authors used Claude (Anthropic) and ChatGPT (OpenAI)  to assist with code debugging and statistical analysis, including the calculation of bootstrap confidence intervals and paired comparisons. All suggested modifications and numerical results were independently reviewed and verified by the authors, who retain full responsibility for the scientific conclusions.


\textbf{Data availability.} The data that support the findings of this article are available from the
author upon reasonable request.


%

=====================================================================
\newpage
\clearpage
\onecolumngrid
\raggedbottom
\setcounter{section}{0}
\setcounter{equation}{0}\setcounter{table}{0}\setcounter{figure}{0}
\renewcommand{\thesection}{S-\Roman{section}}
\renewcommand{\thesubsection}{\thesection.\Alph{subsection}}
\renewcommand{\theequation}{S\arabic{equation}}
\renewcommand{\thetable}{S\arabic{table}}
\renewcommand{\thefigure}{S\arabic{figure}}
\renewcommand{\theHsection}{supp.\arabic{section}}
\renewcommand{\theHsubsection}{supp.\arabic{section}.\arabic{subsection}}
\renewcommand{\theHequation}{supp.\arabic{equation}}
\renewcommand{\theHtable}{supp.\arabic{table}}
\renewcommand{\theHfigure}{supp.\arabic{figure}}

\setcounter{page}{1}
\renewcommand{\thepage}{S\arabic{page}}

\begin{center}
{\large\bfseries Supplemental Material}\\[4pt]
{\large\bfseries Nonlocal multipolar spin accumulation in altermagnets beyond d-wave}\\[4pt]
{J. W. Gonz\'alez}\\[4pt]
Departamento de F\'isica, Universidad de Antofagasta, Av. Angamos 601, Casilla 170, Antofagasta, Chile
\end{center}

\author{J. W. Gonz\'alez}
\affiliation{Departamento de F\'isica, Universidad de Antofagasta, Av. Angamos 601, Casilla 170, Antofagasta, Chile}

\section{Kwant setup and absolute normalization}
The sample is an $N\times N$ square ($d$-, $g$-wave) or a disk of radius 90 sites ($i$-wave, triangular lattice). The disorder $V_j\in[-U/2,U/2]$ acts only inside the region whose distance to the outer edge exceeds the width of the absorbing layer (25 sites for $N=181$, 30 sites for $N=241$). In the absorbing layer, the on-site energy acquires an imaginary part $-i\eta(d)$ with $\eta(d)=0.3\,[(w_{\rm cap}-d)/w_{\rm cap}]^2$, where $d$ is the distance to the edge ($0 \le d \le w_{\rm cap}$). Hermiticity checks are disabled in the solver to accommodate this non-Hermitian term. A three-site chain coupled to the central site with hopping $-1$ is attached to a semi-infinite chain lead. For each spin, we compute the scattering states injected from the lead at $E_F$, the density $n_s(\bm r)=\sum_\alpha|\psi_\alpha(\bm r)|^2$, and divide by $I_s=N_s-R_s$, the number of open lead modes minus the reflection probability. Because Kwant~\cite{Kwant2014} normalizes each incoming mode to unit particle current, $n_s/I_s$ is a density per injected particle current. We verified this normalization in a clean $101\times101$ sample at $E_F=1$ with an absorbing layer of 20 sites: the net outward quantum current through rings of radius 10, 15, 20, and 25 sites, divided by the injected current, is $1.011$, $1.003$, $1.020$, and $0.999$, respectively.

Harmonics are computed on rings of width two sites inside the disordered region, $c_m(r)=2\langle\delta s\,e^{-im\theta}\rangle_{\rm ring}$, and projected onto the symmetry-fixed phase ($\mathrm{Re}\,c_2$ for $d$-, $\mathrm{Im}\,c_4$ for $g$-, $\mathrm{Im}\,c_6$ for $i$-wave). The overall sign is fixed by the clean sample (negative for $d$-wave, so Fig.~2(a) of the main text shows $-\mathrm{Re}\,c_2$). For comparison, the BGK profiles are evaluated as magnitudes $|c_m|$. Only realizations complete at all radii are included, and bootstrap resampling acts on whole realizations.

\section{Boltzmann calculation in bulk and finite geometries}
For each spin, we sample the Fermi contour at $E_F$ and compute the velocity from the gradient of the band energy. We weight the contour average by $d\ell_k/|\bm v_s|$, where $d\ell_k$ is the line element in momentum space. These averages yield the velocity moments and $A_s(\bm q)$. For the bulk reference, we evaluate Eq.~(2) of the main text on a Fourier grid, omit the uniform mode, and inverse-transform the spin-density difference to obtain its real-space profile. For the maps of Fig.~1 of the main text only, a Gaussian low-pass filter of width 1.5 sites removes lattice-scale aliasing of the FFT; it does not affect structure on the scale of $\ell\approx 9$ sites.

Inside a convex region $\Omega$, a free flight between two points of $\Omega$ never leaves $\Omega$. With an absorbing (non-reentrant) boundary, the BGK equation~\cite{BGK1954} is then equivalent to the Peierls equation
\begin{equation}
\rho_s(\bm r)=\sum_{\bm r'\in\Omega}K_s(\bm r-\bm r')\,\rho_s(\bm r')/\tau + K_s(\bm r),\qquad
\hat K_s(\bm q)=\tau A_s(\bm q),
\end{equation}
for $\bm r\in\Omega$, where $\rho_s$ is the density per unit source rate and the discrete sum implicitly includes the unit cell area measure. For $\Omega$ equal to the whole plane, it reduces to Eq.~(2) of the main text. We evaluate the convolution by FFT on a $512\times512$ periodic box, chosen to suppress wrap-around effects, and solve with GMRES to a relative residual below $10^{-10}$. Close to the source, the finite-geometry solution coincides with the infinite-plane reference (e.g., the ratio of finite to infinite amplitudes is $0.986$ for $c_2$ and $1.000$ for $c_4$ at $r=10$ and $U=2$ in the $N=181$ geometry) and departs from it as the boundary is approached. This approximation neglects reflection at the absorbing layer. For the triangular lattice, the BGK density per unit area is multiplied by the cell area $\sqrt{3}/2$ to compare directly with Kwant densities per site. We project the finite-geometry solution onto the same angular harmonics and radial bins used for the Kwant profiles, allowing a direct comparison of magnitudes without amplitude rescaling.

\section{From Fermi-surface moments to real-space multipoles}
The coefficients of Eq.~(3) of the main text are ratios of polynomials in $\hat{\bm q}$; for example
$M_4/M_2^2=\langle(\bm q\cdot\bm v)^4\rangle/\langle(\bm q\cdot\bm v)^2\rangle^2$ is homogeneous of degree zero
and nonanalytic at $q=0$ whenever its angular part is not constant. Expanding $\exp(i\bm q\cdot\bm r)$ in
Bessel functions gives $i^m e^{im\theta}/(2\pi)$ times the radial integral
$\int_0^\infty dq\,q^{p+1}J_m(qr)$. For $m>0$ and $r>0$ this gives
\begin{equation}
\int\frac{d^2q}{(2\pi)^2}\,e^{i\bm q\cdot\bm r}\,q^p e^{im\phi_q}
=\frac{i^m e^{im\theta}}{2\pi}\,\frac{2^{p+1}\,\Gamma\!\left(\frac{m+p+2}{2}\right)}{\Gamma\!\left(\frac{m-p}{2}\right)}\,
\frac{1}{r^{p+2}},
\label{eq:angular-transform}
\end{equation}
where the radial integral is understood in the regularized, distributional sense. For the even $m$ considered here,
the negative angular harmonics follow by complex conjugation. The leading terms for $d$-, $g$-, and
$i$-wave order have $(p,m)=(-2,2)$, $(0,4)$, and $(2,6)$, with transform coefficients $-1/(4\pi)$,
$2/\pi$, and $-96/\pi$ multiplying the corresponding angular harmonic and radial power.

A direct FFT of $q^p e^{im\phi_q}$ with a Gaussian cutoff reproduces the coefficient within 4\% at
$r=50$ for $(0,4)$ and $(2,6)$. The $(0,2)$ case, which avoids the $q=0$ singularity of the physical $(-2,2)$ case, serves as a smooth numerical benchmark and reproduces the coefficient to a relative error of $10^{-5}$ at $r=50$.

For nonnegative even $p$ and even $m\le p$, $q^p e^{im\phi_q}$ is a polynomial in $q_x$ and $q_y$
and produces only terms supported at the source. The $g$-wave term is independent of $q=|\bm q|$
but retains its $\sin 4\phi_q$ angular dependence and produces an $r^{-2}$ tail.

The small-$q$ expansion gives the asymptotic profile for $r\gg\ell$. At $r\sim\ell$ we use the
full solution of Eq.~(2) of the main text.

\section{The \texorpdfstring{$g$}{g}-wave profile at two system sizes}
Table~\ref{tab:plateau} compares the two sample sizes over identical rings ($18.5\le r\le 62.5$ sites,
$2.1\ell$ to $6.9\ell$) and the 150 common disorder realizations. Table~\ref{tab:slopes} lists the local
logarithmic slope of $r^2\,\mathrm{Im}\,c_4$ in successive windows of width $\ell$, with one-standard-deviation bootstrap uncertainties.
Beyond $3\ell$ most local slopes lie within $2\sigma$ of zero, while for $N=241$ the windows
$3$--$4\ell$ and $4$--$5\ell$ deviate by $2.1\sigma$ in opposite directions.
The paired difference in Table~\ref{tab:plateau} measures the change of the window mean over common
rings. Figure~2(b) of the main text shows the outward shift of the downturn in the larger geometry.

The comparison changes both the sample size and the absorbing-layer width, from 25 to 30 sites. The BGK domain $\Omega$ is chosen to match the disordered region of the Kwant setup, terminating at the inner boundary of the absorbing layer, ensuring a direct comparison of the active regions.
\begin{table}[h]
\caption{Mean of $r^2\,\mathrm{Im}\,c_4$ over common rings and realizations ($U=2$, $\ell=9.0$).}
\label{tab:plateau}
\begin{ruledtabular}
\begin{tabular}{lccc}
 & $N=181$ & $N=241$ & paired difference (95\% interval) \\
Kwant & 1.98 & 2.74 & $+0.76$ ($+0.18$ to $+1.31$) \\
BGK (same geometry) & 1.50 & 1.66 & $+0.16$ \\
\end{tabular}
\end{ruledtabular}
\end{table}
\begin{table}[h]
\caption{Local slope of $\log(r^2\,\mathrm{Im}\,c_4)$ versus $\log r$ in successive windows ($U=2$), with one-standard-deviation bootstrap uncertainties.}
\label{tab:slopes}
\begin{ruledtabular}
\begin{tabular}{lcccccc}
window ($r/\ell$) & 1--2 & 2--3 & 3--4 & 4--5 & 5--6 & 6--7 \\
$N=181$ & $+0.99\pm0.37$ & $+0.54\pm0.60$ & $+1.90\pm1.30$ & $-2.64\pm1.37$ & $-1.56\pm1.68$ & $-2.00\pm1.92$ \\
$N=241$ & $+0.68\pm0.28$ & $+0.86\pm0.41$ & $+2.13\pm1.02$ & $-2.95\pm1.43$ & $+0.71\pm1.68$ & $-0.74\pm1.88$ \\
\end{tabular}
\end{ruledtabular}
\end{table}

\section{Disorder scaling}
\label{sec:disorder}
Amplitudes are defined as the absolute values of the projected harmonic, averaged over fixed radial windows and weighted by $r^{m-2}$: 30 to 70 sites for $d$- and $g$-wave order ($N=241$, $U=1.5$, $1.75$, $2.0$, $2.5$; 150 and 300 realizations) and 18 to 40 sites for $i$-wave order ($U=3.0$, $3.5$, $4.0$; 100 common realizations). The exponent $p$ in $A\propto\tau^p$ is obtained via a least-squares fit in $\log\tau$, and its error is derived from a joint bootstrap over the explicit intersection of disorder seeds, which preserves correlations between different $U$ ($10^4$ replicas). Replicas yielding a non-positive window amplitude cannot enter the logarithmic fit and are discarded: none for $d$-wave, 0.01\% for $g$-wave, and 0.8\% for $i$-wave. We obtain $p_d=-1.08\pm0.11$ (95\% interval $-1.30$ to $-0.85$), $p_g=+0.27\pm0.33$ ($-0.18$ to $+1.04$), and $p_i=+1.07\pm0.99$ ($-0.05$ to $+3.66$), against the BGK values in the same windows of $-0.91$, $+0.74$, and $+1.63$. The BGK finite-window exponents differ from the asymptotic values $-1$, $+1$, and $+3$, as the sampled radii are only a few mean free paths from the injector. We exclude $U=3$ for $d$- and $g$-wave order ($k_F\ell\approx 4$), where the Born estimate of $\tau$ becomes less reliable and Kwant departs from BGK. For $i$-wave order, halving the splitting at $U=3.5$ (same realizations) gives a window ratio of $0.70$ (68\% interval $0.52$ to $0.88$), against $0.68$ in BGK.

\section{Independent scattering time}
To test the Born estimate of $\tau$, we compute, for each spin, the disorder-averaged transmission $\langle T\rangle$ of strips of width $W=100$ and length $L=20$ to $140$ sites for the same $d$- and $g$-wave models, using the same box disorder and clean nearest-neighbor leads (60 realizations). The slope of $1/\langle T\rangle$ versus $L$ gives the conductivity per spin, $\sigma=1/(bW)$ in units of $e^2/h$, and the ratio to the Drude value $2\pi\nu\langle v_x^2\rangle_s\tau_{\rm Born}$ yields $\tau_{\rm eff}/\tau_{\rm Born}$; within the Ohmic fit, the lead mismatch contributes primarily to the intercept. We obtain ratios of $0.93$ to $0.95$ at $U=1.5$, $0.85$ to $0.91$ at $U=2$, and $0.79$ to $0.84$ at $U=2.5$ (95\% relative error intervals of $\pm3$ to $6\%$). For $d$-wave order, the two spins yield compatible transport times ($0.85$ and $0.91$ at $U=2$, with overlapping intervals) although their conductivities differ by nearly a factor of two. The ratio decreases from the short to the long half of the $L$ range, consistent with weak localization. We then solve the finite-geometry BGK equation with $\tau=0.88\,\tau_{\rm Born}$ (the lower bound at $U=2$) and $0.93\,\tau_{\rm Born}$ (the upper bound at $U=1.5$) at $U=2$ to test sensitivity. For $d$-wave order, in the window $2\ell<r<8\ell$ of Fig.~2(a) of the main text, the Kwant/BGK ratio changes from $1.15$ (Born) to $1.02$ and $1.07$. For $g$-wave order, over the common rings and 150 common realizations of Table~\ref{tab:plateau}, the ratios change from $1.32$ and $1.65$ ($N=181$ and $241$) to $1.46$ and $1.81$ for $0.88\,\tau_{\rm Born}$, and to $1.40$ and $1.74$ for $0.93\,\tau_{\rm Born}$; with all 300 realizations of $N=241$ and the nominal window $2\ell_{\rm Born}<r<8\ell_{\rm Born}$, the ratio changes from $1.53$ to $1.68$ and $1.61$. Replacing the Born time with the transport time inferred from the strip conductivity reduces the discrepancy for $d$-wave order and increases it for $g$-wave order; this tests a scalar change of $\tau$ only.

To test the boundary, we repeat the $g$-wave calculation for $N=241$, $U=2$, and the 150 common realizations with the strength of the absorbing potential halved ($\eta=0.15$) and doubled ($\eta=0.6$), keeping the disordered region and the disorder realizations fixed. A short run with the reference value $\eta=0.3$ reproduces the original data exactly. Over the common rings of Table~\ref{tab:plateau}, the paired change of $\langle r^2\,\mathrm{Im}\,c_4\rangle$ is $+0.021$ (95\% interval $-0.014$ to $+0.055$) for $\eta=0.15$ and $-0.003$ ($-0.009$ to $+0.003$) for $\eta=0.6$, which is less than 5\% of the gap $2.74-1.52=1.22$ between Kwant and BGK with $0.88\,\tau_{\rm Born}$. At fixed layer width, the absorber strength has little effect on the discrepancy over the range tested. Coherent scattering in the quantum calculation remains a possible contribution.

\section{Two-contact ribbon}
Control ($d_{xy}$, Herasymchuk et al.~\cite{Herasymchuk2025}): nearest-neighbor hopping $-t$, diagonal hoppings $\mp t_m s$ along $(1,\pm1)$, $t_m=0.3t$, $\mu=-2t$, 20\% of the sites with potential $10t_m=3t$; we use $121\times121$ sites and contacts of 11 sites (odd sizes keep the sample symmetric under $y\to-y$) instead of $60\times60$ and 10, and the bias window $0.2t$ sampled with five energies (applied to both the control and $g$-wave models). $g$-wave model: same geometry, $U=2$, $\lambda=1$ and $0.5$. Leads are spin-independent nearest-neighbor metals of the width of the contacts.

The mirror $y\to-y$ maps the sample and the disorder ensemble onto themselves and exchanges the two spin species for the $d_{xy}$ and $g$-wave models; the disorder-averaged spin density is therefore odd in $y$, and the harmonics $\cos 2\theta$ and $\cos 4\theta$ around the injector vanish. In the clean samples the projection on the forbidden parity is zero to machine precision. Harmonics are computed on half rings of width two sites centered on the injecting contact. In each realization, we divide the projected spin density by $n_0$, the spin-summed left-injected density averaged over the same half ring.

Robustness of the $\sin 4\theta$ window amplitude ($7\le r\le 23$ sites, percent of $n_0$): the per-realization distributions have excess kurtosis below $0.6$ (600 realizations), and the mean, 10\% trimmed mean, and Huber estimators agree within errors; the two halves of the first 600 realizations give $-1.56\pm0.38$ and $-1.51\pm0.36$; a $61\times61$ sample with 400 realizations gives $-1.71\pm0.31$; the window was fixed on the first 600 realizations ($-1.53\pm0.25$), and the remaining 1800 independent realizations alone give $-1.04\pm0.15$ ($6.8\sigma$), the held-out validation result quoted in the abstract; the full set of 2400 gives $-1.16\pm0.13$ ($9\sigma$). Disorder dependence (300 to 600 realizations): $-2.0\pm0.3$ at $U=1.5$, $-1.53\pm0.25$ at $U=2.0$ (first 600), and $-0.24\pm0.31$ at $U=2.5$. The paired ratio between $\lambda=0.5$ and $\lambda=1$ over 2400 common realizations is $0.61$ (68\% interval $0.49$ to $0.75$, 95\% interval $0.39$ to $0.90$); in the window $5\le r\le 15$ sites it is $0.56$ (95\% interval $0.34$ to $0.84$).

For the contact-width control we keep the same sample size, $U=2$, $\lambda=1$, and radial window. For widths $w=7,11,15$ sites the signed $\sin 4\theta$ amplitudes are $-1.42\pm0.32\%$, $-1.53\pm0.25\%$, and $-1.40\pm0.20\%$, respectively (evaluated on 600 common realizations for all widths, with 3 additional independent runs for $w=7$ and $w=15$; errors are standard errors of the disorder mean). The corresponding $\sin 2\theta$ amplitudes are $-1.36\pm0.48\%$, $-1.02\pm0.37\%$, and $-0.63\pm0.29\%$. The fourth harmonic is resolved at all three contact widths, with no statistically resolved difference between $w=7$ and $w=15$. The central estimate of the wall-allowed second harmonic becomes less negative as the contacts widen, but the difference is consistent with zero. All three widths use the window fixed at $w=11$ and the same disorder seeds as the first 600 realizations, so this comparison tests the dependence on the contact width and is not an independent estimate of the significance.

\section{Experimental estimates}
We consider detecting the $g$- and $i$-wave signals from an injector narrower than $\ell$ at distances of order $\ell$ to $5\ell$, where the signals are larger than in their far tails. For $\ell=20$ to $50$~nm, these distances span tens to a few hundred nanometers, provided the spin relaxation length exceeds this range. The relative amplitude is controlled by the spin anisotropy of the relevant Fermi-surface moment, which can be of the order of the spin splitting relative to the Fermi energy. In the square geometry of Fig.~2 of the main text, with splittings that are a sizable fraction of $E_F$, the leading harmonic relative to the local charge density at $r=2\ell$ is 8 to 10\% for $d$-wave order and 5 to 8\% for $g$-wave order, in both Kwant and finite-geometry Boltzmann theory. The two-contact ribbon, with the same $g$-wave model but different injection, boundaries, normalization, and energy average, gives about 1\%. The value for a given material depends on its Fermi surface, the contact, and the drain. Nonlocal ferromagnetic detectors~\cite{JohnsonSilsbee1985,Jedema2001} and spin-polarized scanning probes could resolve the spin accumulation. In a reciprocal setup with a spin bias, scanning magnetometry could probe the magnetic field from the induced charge currents \cite{Herasymchuk2025,Ku2020,AharonSteinberg2022}. Relating that field to the spin accumulation requires a device model.

Boltzmann theory predicts that increasing $\tau$ enhances the absolute $g$- and $i$-wave harmonics per injected current and weakens the $d$-wave harmonic, with amplitude proportional to $\tau^{m-3}$. Because $n_0\propto\tau^{-1}$, the relative contrast of Eq.~(5) of the main text scales as $\tau^{m-2}$ at fixed physical distance and band parameters, up to the logarithmic factor. The finite-geometry data resolve the disorder trend only for $d$-wave order (Sec.~\ref{sec:disorder}). For $g$- and $i$-wave order, the mobility dependence remains a prediction that experiments could test. Layered metallic altermagnets such as KV$_2$Se$_2$O and Rb$_{1-\delta}$V$_2$Te$_2$O \cite{Jiang2025KV2Se2O,Zhang2025RbV2Te2O} are promising candidate $d$-wave platforms; for $g$-wave materials the film orientation must preserve the in-plane fourfold harmonic of the splitting.

\end{document}